\documentclass[a4paper,11pt]{article}
\usepackage{jcappub} 
\usepackage{lineno}

\title{\boldmath Hubble tension in a nonminimally coupled curvature-matter gravity model}

\author[a]{Miguel Barroso Varela}

 \affiliation[a]{Departamento de Física e Astronomia, Faculdade de Ciências, Universidade do Porto,\\Rua do Campo Alegre s/n, 4169-007 Porto, Portugal}

\author[a,b]{and Orfeu Bertolami}%

 \affiliation[b]{Centro de Física das Universidades do Minho e do Porto,\\Rua do Campo Alegre s/n, 4169-007 Porto, Portugal}%

\emailAdd{up201907272@edu.fc.up.pt}
\emailAdd{orfeu.bertolami@fc.up.pt}

\abstract{The presently open problem of the Hubble tension is shown to be removed in the context of a modified theory of gravity with a non-minimal coupling between curvature and matter. By evolving the cosmological parameters that match the cosmic microwave background data until their values from direct late-time measurements, we obtain an agreement between different experimental methods without disrupting their individual validity. These modified gravity models are shown to provide adequate fits for other observational data from recent astrophysical surveys and to reproduce the late-time accelerated expansion of the Universe without the inclusion of a cosmological constant. This compatibility with observations presents further evidence of the versatility of these models in mimicking diverse cosmological phenomena in a unified manner.}

\begin{document}
\maketitle
\flushbottom

\section{Introduction}
Cosmology can be a highly volatile field of study, in great part thanks to the rapidly evolving set of experiments and their increasing accuracy. Measurements dating back to the late 1920s have led to the conclusion that we live in an expanding Universe. Naturally, the rate of such an expansion has been a crucial measurement for many decades, starting with Hubble's first discovery in 1929 \cite{HubbleOriginal}. The rate is typically described through the Hubble parameter $H=\dot a /a$, where $a$ is the scale factor of the Universe and a dot represents a derivative with respect to cosmic time. As the accuracy and diversity of measurements of the present value of the Hubble parameter ($H_0$) increased, it became clear that there is a significant tension between so-called early and late experimental methods.\par
Late measurements of the Hubble parameter are those which utilise direct observations of cosmological effects at low redshifts to infer $H_0$. The purely empirical method uses the distance-redshift relation, which is implemented through the construction of a ``distance ladder". Cepheids and type Ia supernovae are particularly useful for this purpose, as their large and theoretically predictable luminosity allows for accurate observations over several Megaparsecs (Mpc). \par
The Hubble Space Telescope (HST) was the first experiment to provide the means to measure these objects at sufficiently large distances to yield a value of $H_0=72\pm 8 \ \text{km} \ \text{s}^{-1} \ \text{Mpc}^{-1}$ \cite{Hubble2001}. Recent improvements in the accuracy of these distance estimates have resulted in values in the range of 73-74 $\text{km} \ \text{s}^{-1} \ \text{Mpc}^{-1}$, with the most recent set of 75 Milky Way Cepheids with HST photometry and EDR3 parallaxes \cite{RecentLateMeasurements} yielding $H_0=73.2\pm1.3 \ \text{km} \ \text{s}^{-1} \ \text{Mpc}^{-1}$ \cite{R20Cepheids}, which will be taken as the reference value for late measurements throughout this paper.\par
Early measurements of the Hubble parameter are commonly considered as those relying on observations at redshifts $z>1000$ and assuming agreement with the $\Lambda$CDM model at those times \cite{HubbleTensionReview}. The same model is then used to evolve the Universe until $z=0$, thus obtaining estimates for cosmological parameters in the present. The Planck experiment is widely taken to be the standard for early measurements, predicting $H_0=67.27\pm0.60 \ \text{km} \ \text{s}^{-1} \ \text{Mpc}^{-1}$ in a flat $\Lambda$CDM model from observations of the cosmic microwave background (CMB) radiation in their 2018 results \cite{Planck2018}. Additionally, measurements of Baryon Acoustic Oscillations (BAO) with a CMB prior give a similar value of $H_0=67.9\pm1.1 \ \text{km} \ \text{s}^{-1} \ \text{Mpc}^{-1}$ \cite{BAO_BOSSData}. Late and early methods for the determination of $H_0$ are thus in a tension of several standard deviations which has yet to be definitively resolved \cite{HubbleTensionReview}. \par 
Several solutions have been proposed to relieve or even remove the Hubble tension, with possibilities ranging from suggestions of possible systematic errors to modifications of the standard cosmological model \cite{HubbleTensionReview}. The latter option is typically based on altering the evolution from CMB data to the present by adapting the underlying gravity theory in some way. Such modifications include postulating the presence of accelerated expansion in the early Universe as described in early dark energy models \cite{EDE}, late dark energy models like those with a time-varying equation of state parameter \cite{LateDE}, Über-gravity models \cite{uberGravity}, Galileon gravity models \cite{GalileonGravity}, among many others (see Refs. \cite{HubbleTensionReview,HubbleTensionReview2,HubbleTensionReview3} for extensive reviews of solutions). Relevantly for the method considered in this paper, different models of modified gravity have been suggested as possible solutions to the current tension. These include $f(R)$ \cite{F(R)HubbleTension,F(R)HubbleTension2}, $f(T)$ \cite{F(T)HubbleTension,F(T)HubbleTension2,F(R)HubbleTension3} and $f(Q)$ \cite{F(Q)HubbleTension} models, along with different scalar-tensor theories \cite{ScalarTensorHubbleTension,ScalarTensorHubbleTension2}. These proposals provide varying degrees of success, with some only alleviating the $4\sigma$ tension to roughly $(2-3)\sigma$ and others providing means to fully resolve the gap between different experiments \cite{HubbleTensionReview}. The landscape of solutions shows greater success for modifications of the late-time Universe \cite{HubbleTensionReview2}. \par
On top of this tension, there is also very active research on the topic of the accelerated expansion of the Universe, which is not only observed to be expanding but also seems to be doing it increasingly faster as time progresses. While different methods for fitting the observational data have been used to account for this hypothesis \cite{AcceleratedExpansion,AcceleratedExpansion2,AcceleratedExpansion3}, there is still an active debate over what could be causing this acceleration. Clearly, the simplest proposal is the inclusion of a cosmological constant $\Lambda$, which raises unanswered questions about its seemingly fine-tuned magnitude and the associated matter/dark energy transition time, while also being in disagreement with predictions made by quantum field theory arguments \cite{CosmologicalConstantProblem}. Alternative proposals on these issues have been presented in Refs. \cite{NMCAcceleratedExpansion,NMCCosmologicalConstant,InflationCosmologicalConstant,ChaplyginGasVacuum}. \par
In this work, we consider a modification of the expansion rate of the Universe between early and late measurements caused by a modified theory of gravity with non-minimal coupling (NMC) of matter and curvature \cite{ExtraForce}. This is similar to the work conducted in Refs. \cite{F(R)HubbleTension,F(R)HubbleTension2}, where only minimally coupled $f(R)$ theories were considered. Additionally, we seek solutions which simultaneously display accelerated expansion at low redshifts without the inclusion of a cosmological constant, as investigated in Refs. \cite{NMCAcceleratedExpansion,NMCCosmologicalConstant}. Such theories have also been extensively researched in the context of mimicking dark matter profiles \cite{NMCDarkMatter,NMCDarkMatter2}, analysing the modified theory with solar system constraints \cite{NMCSolarSystem,NMCSolarSystem2,NMCSolarSystem3}, sourcing cosmological inflation in the early Universe \cite{NMCInflation,NMCInflation2,NMCInflation3} and the creation of large-scale structure \cite{NMCCosmologicalPerturbations}.\par
The layout of this paper is as follows. We present the nonminimally coupled model, the relevant field equations and their implications for the Friedmann-Lemaître-Robertson-Walker (FLRW) metric in Section \ref{NMCSection}. The methodology for relieving the Hubble tension, the necessary numerical methods and the obtained results are described in Section \ref{MethodSection}. This is followed by a discussion of matching late-time acceleration effects in NMC models with the resolved Hubble tension and their comparison with empirical fitting methods, which is included in Section \ref{AccelerationSection}. We conclude the paper in Section \ref{ConclusionSection}, where we discuss the obtained results along with possible extensions of our work.
We use the $(-,+,+,+)$ signature, choose units where $c=1$ and define $8\pi G=\kappa^2$.

\section{Nonminimally coupled model}\label{NMCSection}
\subsection{Action and field equations}
The nonminimally coupled $f(R)$ model can be written in action form as \cite{ExtraForce}
\begin{equation}
    S=\int dx^4 \sqrt{-g} \left[\frac{1}{2\kappa^2}f_1(R)+[1+f_2(R)]\mathcal{L}_m \right],
\end{equation}
where $f_{1,2}(R)$ are arbitrary functions of the scalar curvature $R$, $g$ is the metric determinant and $\mathcal{L}_m$ is the Lagrangian density for matter fields \cite{ExtraForce}. The General Relativistic action is recovered by setting $f_1=R$ and $f_2=0$. The inclusion of a cosmological constant can be considered by choosing $f_1=R-2\kappa^2 \Lambda$. By varying the action with respect to the metric $g_{\mu\nu}$ we obtain the field equations \cite{ExtraForce}
\begin{equation}\label{FieldEquations}
    (F_1+2\kappa^2 F_2\mathcal{L}_m)G_{\mu\nu}=\kappa^2(1+f_2)T_{\mu\nu}+\Delta_{\mu\nu}(F_1+2\kappa^2 F_2 \mathcal{L}_m)+\frac{1}{2}g_{\mu\nu}(f_1-F_1 R - 2\kappa^2 F_2 R \mathcal{L}_m),
\end{equation}
where we have defined $\Delta_{\mu\nu}\equiv\nabla_\mu\nabla_\nu-g_{\mu\nu}\Box$ and $F_i\equiv df_i/dR$. Taking the covariant derivative of both sides of Eq. (\ref{FieldEquations}) and using the Bianchi identities $\nabla_\mu G^{\mu\nu}=0$ leads to the non-conservation law \cite{ExtraForce}
\begin{equation}\label{NonConservationEq}
    \nabla_\mu T^{\mu\nu}=\frac{F_2}{1+f_2}\left(g^{\mu\nu}\mathcal{L}_m-T^{\mu\nu}\right)\nabla_\mu R,
\end{equation}
which reduces to the usual stress-energy tensor conservation for $f_2=0$. This non-conservation follows directly from the non-minimal coupling of matter and curvature. In this work, as we aim to determine the effects of the NMC model independently of the minimally coupled $f(R)$ model, we set $f_1=R$ and consider $f_2\neq0$.

\subsection{Cosmology in nonminimally coupled gravity}
Next, we use the flat FLRW metric given by the line element
\begin{equation}
    ds^2=-dt^2+a^2(t)\left(dr^2+r^2 d\Omega^2\right)
\end{equation}
and the usual perfect fluid stress-energy tensor components $T_{00}=\rho$ and $T_{rr}=a^2 p$, where $\rho$ and $p$ are the energy density and the isotropic pressure of the fluid, respectively. As can be seen from the field and non-conservation equations, the choice of $\mathcal{L}_m$ is non-trivial, as it explicitly enters these equations via the nonmiminal coupling, while only appearing in the form of the related stress-energy tensor in General Relativity \cite{LagrangianChoice,LagrangianForm}. With this in mind, in the remainder of this work, we follow the arguments given in Refs. \cite{LagrangianForm,LagrangianChoice2} and take the Lagrangian density to be $\mathcal{L}_m=-\rho$. \par
We can then use the previously discussed field equations (\ref{FieldEquations}) to arrive at the modified Friedmann equation
\begin{equation}\label{ModifiedFriedmann}
    H^2=\frac{1}{6F}\left[2(1+f_2)\tilde\rho-6H\dot F-f_1+F R\right],
\end{equation}
where we have defined $\tilde\rho\equiv\kappa^2\rho$ and $F\equiv F_1+2\kappa^2 F_2 \mathcal{L}_m=F_1-2F_2\tilde\rho$ for simplicity \cite{NMCFriedmann}. Additional components of the field equations yield the modified Raychaudhuri equation
\begin{equation}\label{ModifiedRaychaudhuri}
    2\dot H+3 H^2=-\frac{1}{2F}\left[2 \ddot F+4H\dot F+f_1-F R+2\kappa^2(1+f_2)p\right]
\end{equation}
and the non-conservation equation leads to the usual result
\begin{equation}\label{ConservationEq}
    \dot \rho +3H(\rho+p)=0,
\end{equation}
where the choice of $\mathcal{L}_m=-\rho$ causes the modifications in Eq. (\ref{NonConservationEq}) to vanish. This greatly simplifies the cosmological evolution of our system, as we can consider the usual evolution of the energy density of all kinds of matter with respect to the scale factor $\rho\propto a^{-3(1+\omega)}$, which depends only on the equation of state parameter $\omega=p/\rho$ ($\omega=0$ for non-relativistic matter, $\omega=1/3$ for radiation).  \par
For the analysis considered here, it is convenient to explicitly write the deceleration parameter 
\begin{equation}
    q=-\frac{\ddot a a}{\dot a^2}=1-\frac{R}{6H^2}
\end{equation}
and use Eqs. (\ref{ModifiedFriedmann}) and (\ref{ModifiedRaychaudhuri}) to write 
\begin{equation}\label{qEquation}
    H^2-\frac{R}{3}=\frac{1}{F}\left[F_2\tilde\rho R+\ddot F+2H\dot F+\kappa^2 f_2 p\right]+\kappa^2 p,
\end{equation}
where given our choice $f_1=R$, we have $F=1-2F_2\tilde\rho$. This gives $q=\frac{1}{2}+\frac{\kappa^2 p}{2H^2}$ when setting $f_2=0$ and assuming a barytropic equation of state-dominated Universe, as expected from General Relativity. More importantly, it correctly implies that $q=1/2$ when $f_2=0$ and $p=0$, which is the expected result in a matter-dominated Universe, such as the one that we consider as the starting point of the modified regime. This second-order differential equation is useful when evolving the system numerically, as seen in Section \ref{MethodSection}. \par
Throughout the rest of this paper, we will restrict our analysis to the aforementioned background quantities in the modified Friedmann equation. However, one could also consider the effects of this theory on cosmological perturbations, which lead to the creation of large-scale structures in the Universe. This has been studied in detail in Ref. \cite{NMCCosmologicalPerturbations}, where it was shown that the nonminimal coupling can satisfy the weak lensing condition, as well as the necessary frictional attenuation in the growth of density perturbations, as long as the relevant term in $f_2$ has a negative power of $R$, such as the one considered in our work. We shall thus focus on the problem of the Hubble tension and the accelerated expansion of the Universe, keeping in mind that the same mechanism satisfies the necessary conditions for sensible cosmological perturbations and the ensuing formation of large-scale structures. 

\section{Hubble Tension}\label{MethodSection}
In order to ensure the compatibility of late and early measurements, we assume that the CMB data is accurate and that it is the evolution from early to late times that causes the discrepancy between the different values for $H_0$ in the present. As will be discussed in the following section, we assume a decoupling of matter and curvature at high redshifts, allowing us to consider starting conditions compatible with General Relativity
\begin{equation}\label{InitialConditions}
    H^2(z_i)=\frac{\tilde\rho(z_i)}{3}=H_0^{*2}\Omega_m(z_i) \quad \quad \quad  R(z_i)=\tilde\rho(z_i),
\end{equation}
where we have kept only the non-relativistic matter energy density, as the radiation density is negligible at points where the curvature is small enough to ensure a noticeable effect from $f_2$. We also assume no cosmological constant, as the accelerated expansion will be generated by the non-minimal coupling modification. The values $H_0^*$ and $\tilde\rho(z_i)$ are taken from the Planck data \cite{Planck2018}, as this would be the measured value in the present if there were no modifications to gravity and the Universe evolved with a cosmological constant, which is not relevant for the CMB observations due to the small contribution of dark energy at these redshifts. This agreement is also ensured by the standard $\tilde\rho\propto a^{-3}\propto(1+z)^3$ dependence remaining unaltered with our choice of the Lagrangian density. By evolving the Universe from an initial point between early and late measurements, we can obtain a modified present value $H_0$ that is in agreement with direct observations conducted at smaller redshifts. This resolves the tension between experiments with no major implications for the independent measurements.

\subsection{Choice of \texorpdfstring{$f_2(R)$}{Lg}}
A reasonable assumption for cosmology at early times is that $f_2(R)\rightarrow0$ as $R\rightarrow\infty$, as this ensures a decoupling of matter and curvature at high redshifts, such as the ones considered for the CMB measurements and any radiation dominated era. This was already considered in the context of mimicking the effects of dark matter \cite{NMCDarkMatter} and the accelerated expansion of the Universe \cite{NMCAcceleratedExpansion}. A natural choice is then any inverse power of $R$ such as 
\begin{equation}
    f_2(R)=\left(\frac{R_0}{R}\right)^n,
\end{equation}
which we can think of as an isolated term in a more complex expansion of integer powers of $n$. The constant $R_0$ serves as a characteristic curvature for the modification, which can be interpreted as a characteristic length scale $r_c=R_0^{-1/2}$ with units of distance. \par
Naturally, such a choice of $f_2$ leads to the possibility of diverging behaviour when the scalar curvature vanishes \cite{NMCNuclear}. If one thinks of systems with $R=0$, such as the vicinity of an uncharged black hole, the issue of a diverging $f_2$ term seems evident. However, this argument fails to capture the transition of the locally sourced values of the curvature onto its cosmologically sourced behaviour, which becomes relevant when considering large enough scales. Additionally, as discussed in Ref. \cite{NMCAnalitic}, such NMC models are not to be thought of as being composed of functions with unique powers of the curvature scalar, but as a sum of terms that would be relevant at different scales. In the example of nuclear physics given in Ref. \cite{NMCNuclear}, this would simply mean that at these scales the NMC effects are suppressed. Details on the effects of non-analytic forms of $f_2$ in the context of the Solar System were presented in Refs. \cite{NMCSolarSystem3,NMCAnalitic}, where a screening mechanism that allows for planetary constraints to be calculated was introduced. Further discussions of the consistency of these types of inverse $R$ dependence can be found in Ref. \cite{NMCViability}, where stability and causality tests of these theories were addressed. \par
Treating the full form of $f_2$ as a series with integer powers of $R$ has non-trivial consequences on our discussion of decoupling at CMB times. Considering the typical matter-dominated CMB redshift of $z_{\text{CMB}}\simeq1100$ \cite{Planck2018} and the increase of $R$ with density, one should think of the possible emergence of positive powers of $R$, which could break our initial assumptions about the decoupling of curvature and matter. However, at this epoch, we expect the curvature to still not be significant enough to produce sizeable effects on the relevant physics via positive power terms, although this is no longer the case when considering much larger redshifts, such as those associated with the inflationary epoch, where these positive powers of $R$ become relevant and even provide a mechanism for inflation \cite{NMCInflation,NMCInflation2,NMCInflation3}. With this in mind, we take the NMC effects near the CMB epoch to be negligible and thus the modified theory should not have an effect on the CMB and related observables.

\subsection{Numerical Method}
Due to the degree of complexity of the resulting equations, the evolution of the scale factor and other cosmological parameters is conducted via a numerical analysis. While the modified Friedmann equation is an obvious candidate for describing the dynamics of the system, it is found to be prone to numerical instabilities. We thus evolve our system using the second-order Eq. (\ref{qEquation}), which can be usefully rewritten using the relation $d(\log a) =H dt$ as 
\begin{equation}
    \frac{d^2 \left(F_2 \tilde \rho\right)}{d(\log a)^2}=-\frac{1}{2}(1-2F_2 \tilde\rho)\left(1-\frac{R}{3 H^2}\right)-\frac{R}{6H^2}\frac{d\left(F_2\tilde\rho\right)}{d(\log a)}+F_2\tilde\rho\frac{R}{2H^2},
\end{equation}
where we have combined several terms such that we evolve the quantity $F_2 \tilde\rho$ as a proxy for $R$ throughout the simulation. Due to our simple choice of $f_2$, this function can be easily inverted to obtain $R$ at each point of the numerical evolution. We evolve $H$ using the unaltered equation
\begin{equation}
    \dot H=\frac{R}{6}-\frac{H^2}{3}\Rightarrow\frac{dH}{d(\log a)}=\frac{R}{6H}-\frac{H}{3},
\end{equation}
which, together with the previous equation, sets up the dynamics of our system with $\log a$ as our ``time" parameter. The evolution of the matter density is directly determined by its relation with $a=e^{\log a}$. We then take the usual $\Lambda$CDM initial conditions 
\begin{equation}
    H^2_i=\frac{\tilde\rho_i}{3}\quad\quad R_i=\tilde\rho_i ,
\end{equation}
where $\tilde\rho_i=\tilde\rho(z_i)$ is the matter energy density evaluated at the initial redshift, as seen in Eq. (\ref{InitialConditions}). The remaining condition is 
\begin{equation}
\begin{aligned}
    \frac{d\left(F_2\tilde\rho\right)}{d(\log a)}=\frac{1}{H_i}\frac{d(F_2\tilde \rho)}{dt}\bigg\rvert_{z=z_i}=&\frac{1}{H_i}\left(\dot F_2 \tilde\rho_i+F_2\dot{\tilde\rho}_i\right)=\frac{1}{H_i}\left(F_2'\dot R_i \tilde \rho_i-3 H_i F_2 \tilde\rho_i\right)=\\
    =&\frac{\dot{\tilde\rho}_i}{H_i}F_2'\tilde\rho_i-3 F_2\tilde\rho_i=-3F_2'\tilde\rho^2_i-3F_2\tilde\rho_i,
    \end{aligned}
\end{equation}
where we have taken $F_2$ and its derivative $F_2'$ to be evaluated at the initial point with $R_i=\tilde\rho_i$. Thus, the 3 quantities we evolve are $H$, $F_2\tilde\rho$ and its derivative with respect to $\log a$. \par 
The starting point is determined by ensuring an adequately small value of the quantity $F_2 \tilde\rho$ that allows a smooth transition between the standard and modified regime. For a given $n$ and some arbitrary small parameter $\epsilon$ this is
\begin{equation}
    [F_2 \tilde\rho]_{z=z_i}=\epsilon\Rightarrow z_i=-1+\left(\left(\frac{n}{\epsilon}\right)^{1/n}\frac{R_0}{\tilde\rho_0}\right)^{1/3}, 
\end{equation}
where $\tilde\rho_0$ is the value of the energy density at $z=0$ taken from the Planck data \cite{Planck2018}. We can safely consider this value due to the conservation Eq. (\ref{ConservationEq}) being the same as in General Relativity for the NMC theory with the chosen form of the matter Lagrangian density. \par
The determination of the required $R_0$ values to solve the tension is based on defining starting values for $H_0^{\text{direct}}$ and $H_0^{\text{indirect}}$, with the former being taken from Ref. \cite{R20Cepheids} and serving as a target for the simulation, while the latter is taken from Planck's data \cite{Planck2018} and is used to generate the initial conditions of the numerical method. For fixed $n$, each $R_0$ then corresponds to some final $H$ value at $z=0$, which can be used as the input function in a root-finding method, together with the directly observed value of $H_0$ and initial choices of $R_0$ taken from Ref. \cite{NMCAcceleratedExpansion}. We find that the simulated $H_0$ is monotonic with respect to $R_0$, thus allowing us to uniquely determine the necessary form of NMC that eliminates the tension.

\subsection{Results}
\begin{figure}[h!]
    \centering
    \includegraphics[width=0.55\linewidth]{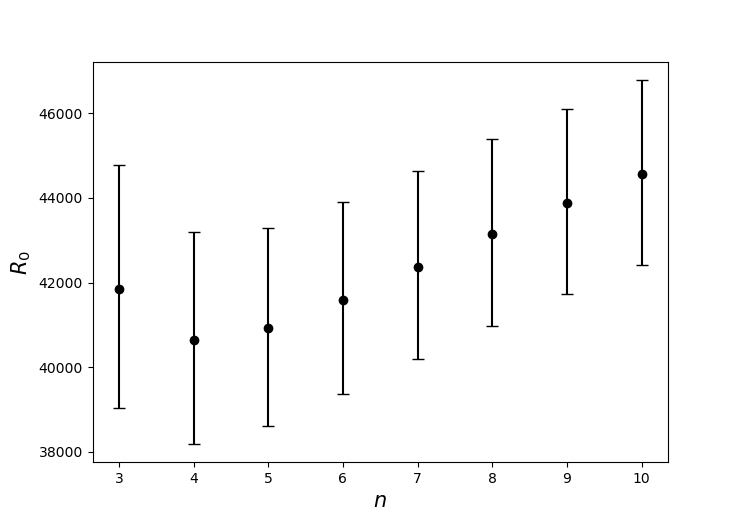}
    \caption{Values of $R_0$ required to remove the Hubble tension for different powers $n$ of the curvature scalar. The y-axis is in arbitrary units and simply serves to provide a comparison between magnitudes of values.}
    \label{R0Plot}
\end{figure}

The obtained values of $R_0$ are shown in Figure \ref{R0Plot}. Results for $n=1,2$ are not presented, as these powers do not enforce a significant modification of the evolution of the Hubble parameter, thus being unable to resolve the tension. Non-integer values of $n$ have similar effects to those shown in the remainder of this work, with the relatively smooth form of the relation between $R_0$ and $n$ allowing for a simple extrapolation for non-integer exponents. The error on the determined values of $R_0$ for each $n$ has been estimated by considering the effect of varying both the late and early measurements of the Hubble parameter and corresponding matter density. A larger value of a late-time determination of $H_0$ leads to the largest possible required $R_0$ to remove the tension for a fixed early-time value. Conversely, larger values of indirect measurements of $H_0$ reduce the required $R_0$ for fixed late-time values. This allows us to easily combine the errors of both ends of the data range and obtain an overall error estimate for each $R_0$. The associated characteristic lengths $r_c=R_0^{-1/2}$ are all around 1450 Mpc, which is comparable to the Hubble length $r_H\approx4000 \ \text{Mpc}$. This illustrates the scale at which the effects of these modifications to the theory would become significant. \par 

For the presented values, it is interesting to note that $n=4$ seems to be at a relative minimum, while the necessary values of $R_0$ seem to increase for larger values of $n$. This follows from the late impact of these terms on the evolution of the Universe, which then requires a larger value of $R_0$ to reconcile the evolution of $H(z)$ with the late-time observational value of $H_0$. All of these values would be smaller if we had included a cosmological constant, as it would then be a matter of patching the approximately $10\%$ difference between values of $H_0$. However, as will be discussed in Section \ref{AccelerationSection}, the absence of a cosmological constant can be accounted for by the effects of the NMC model. With this in mind, we have taken the more encompassing approach, which not only attempts to provide possible solutions to the Hubble tension but also the accelerated expansion problem.

\subsection{Comparison with observational data}
Besides analysing its present value, we can also compare the modified evolution of the Hubble parameter with recent observational data from Cosmic Chronometers (CCs) and Baryon Acoustic Oscillations (BAO). Data for the former were taken from Refs. \cite{CC1,CC2,CC3,CC4,CC5,CC7}, while the latter were taken from Refs. \cite{BAO1,BAO2,BAO3,BAO4,BAO5,BAO6,BAO7,BAO9,BAO10,BAO11}. For this comparison, we consider the $n=4$ and $n=10$ model simulations, as these were found to encapsulate the general behaviour of smaller and larger values of $n$ while being within computationally reasonable boundaries \cite{NMCAcceleratedExpansion}. We compare our models with theoretical expectations from the $\Lambda$CDM model assumed by the Planck experiment \cite{Planck2018}.\par

\begin{figure}[h!]
\centering
    \includegraphics[width=0.55\linewidth]{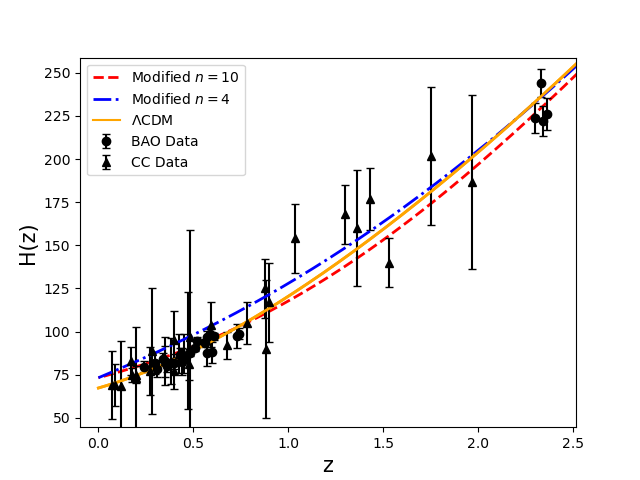}
    \caption{Comparison of modified model with various Baryon Acoustic Oscillation and Cosmic Chronometer data. The flat $\Lambda$CDM model with 2018 Planck data \cite{Planck2018} is shown for the standard theory.}
    \label{CCBAOData}
\end{figure}

The results are shown in Figure \ref{CCBAOData}. The smaller $z$ data points seem to be in better agreement with the $\Lambda$CDM model, while some of the larger $z$ BAO data are surprisingly more in line with the $n=10$ modified model, which has a noticeably better fit than the $n=4$ model. This is confirmed by the $\chi^2$ test of the quality of each fit - $\chi^2_{4}=26$, $\chi^2_{10}=18$ and $\chi^2_{\Lambda\text{CDM}}=15$. However, as discussed at the start of this paper, the BAO values are obtained using a prior from $\Lambda$CDM and are therefore naturally inclined to agree with its predictions. Additionally, as noted in Ref. \cite{HubbleTensionReview}, the CC values have large uncertainties and therefore provide relatively poor insight into late values of the Hubble parameter. Nevertheless, the similarity between $\Lambda$CDM evolution and the considered NMC models with no cosmological constant is already a relevant result, as the standard matter-dominated Universe would fall significantly short of data points at small values of $z$. \par

\begin{figure}[h!]
    \centering
    \includegraphics[width=0.49\linewidth]{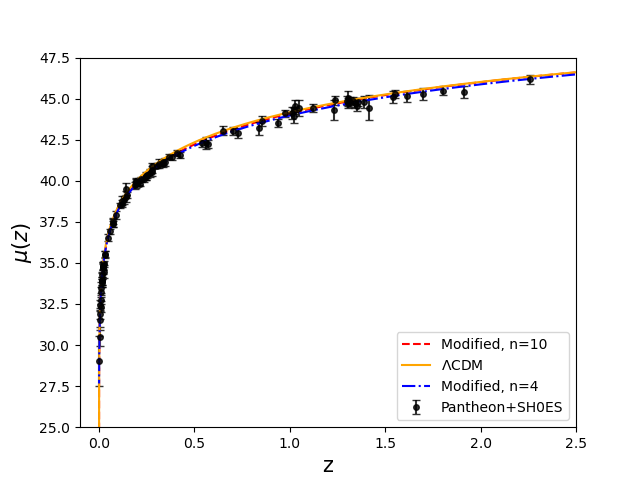}
    \includegraphics[width=0.49\linewidth]{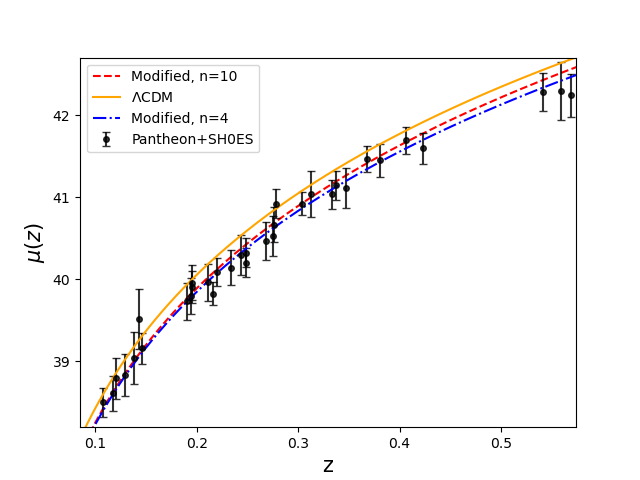}
    \caption{Comparison of modified model with Pantheon+SH0ES distance moduli data \cite{PantheonData,SHOESData}. To allow for better visualisation, points in dense data regions have been removed for this figure. The flat $\Lambda$CDM model with 2018 Planck data \cite{Planck2018} is shown for comparison.}
    \label{DistanceModuli}
\end{figure}

Additionally, we can use the simulated $H(z)$ data to calculate the theoretical distance moduli of astronomical objects at redshift $z$ as
\begin{equation}
    \mu(z)=5\log_{10}\left(\frac{d_L(z)}{1 \text{Mpc}}\right),
\end{equation}
where the luminosity distance is given by 
\begin{equation}
    d_L(z)=c(1+z)\int^z_0\frac{dz'}{H(z')},
\end{equation}
allowing us to add yet another comparison with observational data. The Pantheon+SH0ES dataset contains 1701 light curves of 1550 supernovae in the range $0.001\leq z\leq2.2613$ from 18 surveys \cite{SHOESData,PantheonData}. These measurements are model-independent, thus providing a solid comparison for our results. These are shown in Figure \ref{DistanceModuli}. Due to the large number of data points, it becomes difficult to assess the quality of the fit of each model, leading us to the comparison of their respective $\chi^2$ values. These are $\chi^2_{4}=875$, $\chi^2_{10}=990$ and $\chi^2_{\Lambda\text{CDM}}=2361$. This expresses quantitatively the better fit provided by modifying the evolution of CMB data with the NMC theory. Even when considering a $\Lambda$CDM model with late values of the Hubble constant, we obtain $\chi^2_{\Lambda\text{CDM}}=902$, which is at the quality level of the NMC theory fit. However, this value of $H_0$ would fail to reproduce the CMB data observed by Planck \cite{Planck2018}, while the NMC model can accurately match both ends of the observations. Therefore, we conclude that these modified models go beyond the initially proposed task of matching present values of the Hubble parameter by providing an alternative method to fit existing observational data.

\subsection{Fitting model to observational data}
\begin{figure}
    \centering
    \includegraphics[width=0.75\linewidth]{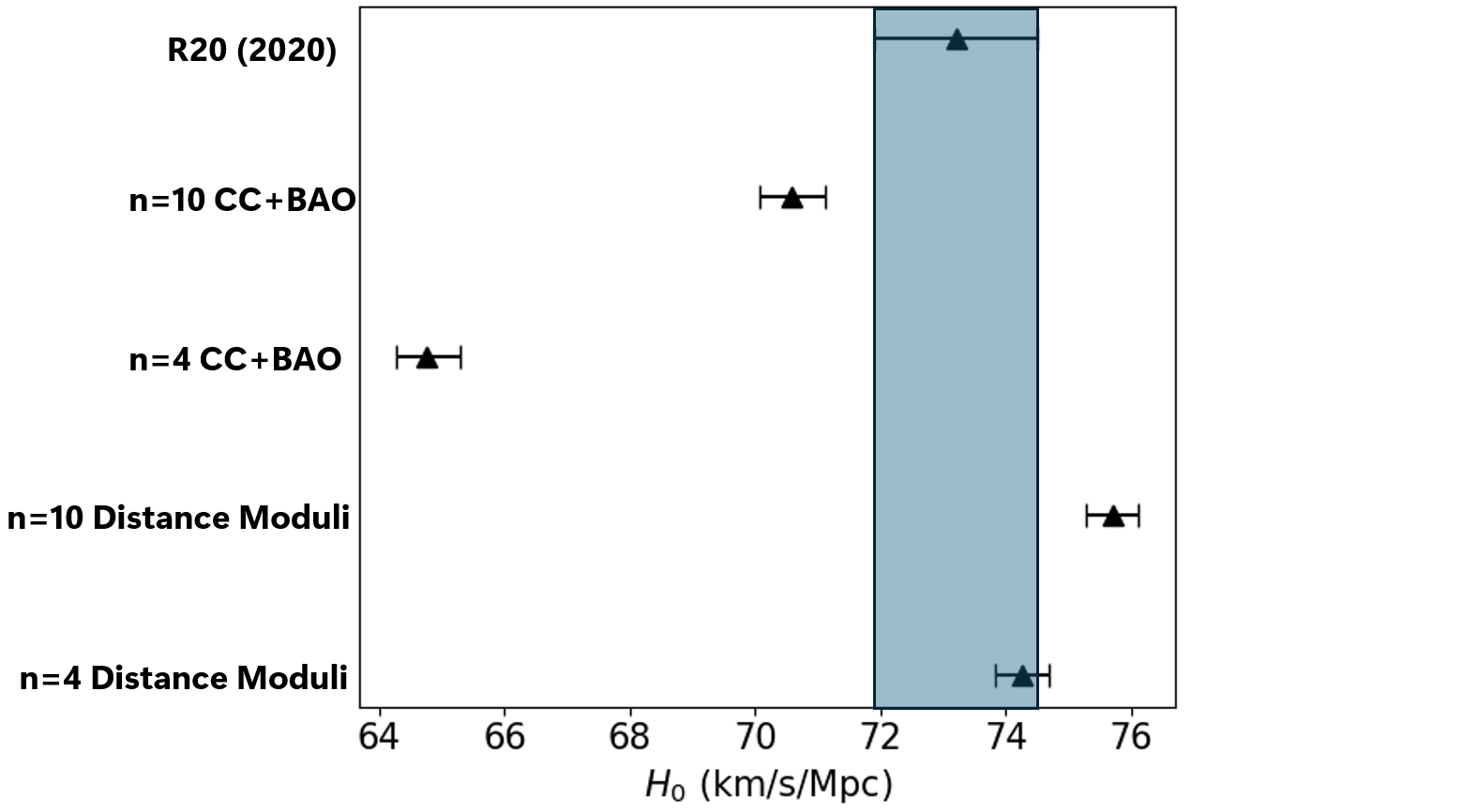}
    \caption{Predicted $H_0$ values in the NMC modified model after fitting to Pantheon+SH0ES collaboration data for distance moduli and CC+BAO data for $H(z)$ separately. The R20 reference value from Ref. \cite{R20Cepheids} is shown, with its standard deviation extended throughout the plot for improved visualisation. }\label{H0Fit}
\end{figure}

Apart from directly matching the early and late measurement methods for $H_0$ and then comparing the same modified model dynamics to observational data, we can also fit the model's parameters to those observations and then present the expected $H_0$ value that follows from the ensuing evolution of the NMC theory. The fitting is conducted by minimising the $\chi^2$ value (or equivalently maximising the likelihood $L\sim e^{-\chi^2}$) by variation of the free parameter $R_0$. From this analysis, we can also extract the standard deviation for $R_0$, leading to a final prediction for a mean and standard deviation of $H_0$. Due to the much larger number of points and increased accuracy provided by the Pantheon+SH0ES measurements of distance moduli \cite{PantheonData,SHOESData}, we fit this data separately from the CC+BAO data used in the previous discussion. This also serves as a useful separation of measurements that present individual values of $H(z)$ at specific redshifts, as is the case for CC+BAO \cite{BAO_BOSSData,CosmicChronometers}, and others that provide quantities that are directly related to the late-time evolution of $H(z)$ as a whole, such as the distance moduli $\mu(z)$ \cite{R20Cepheids}. An overview of the results is shown in Fig. \ref{H0Fit}. \par

By fitting the NMC model to the Pantheon+SH0ES distance moduli data, we obtain $H_0=74.27^{+0.43}_{-0.45}\ \text{km} \ \text{s}^{-1} \ \text{Mpc}^{-1}$ for $n=4$ and $H_0=75.72^{+0.40}_{-0.45}\ \text{km} \ \text{s}^{-1} \ \text{Mpc}^{-1}$ for $n=10$. Comparing with the reference value for late-time measurements of $H_0=73.2\pm1.3\ \text{km} \ \text{s}^{-1} \ \text{Mpc}^{-1}$ \cite{R20Cepheids}, we see that the fitted $n=4$ modified dynamics predict a value for the Hubble constant within error of the direct model-independent measurements. Hence, it is possible to state that the NMC model is able to evolve the Universe from the CMB data in such a way that agrees with the direct late-time observations, thus removing the tension. \par
By fitting the NMC model to the combined CC+BAO $H(z)$ data, we obtain $H_0=64.75^{+0.53}_{-0.47}\ \text{km} \ \text{s}^{-1} \ \text{Mpc}^{-1}$ for $n=4$ and $H_0=70.59^{+0.52}_{-0.53}\ \text{km} \ \text{s}^{-1} \ \text{Mpc}^{-1}$ for $n=10$. Again comparing to the same reference value as above, we see that both of the fitted values lie significantly below it. However, this follows from the expected bias towards Planck+$\Lambda$CDM predictions present in the BAO data, as discussed in Ref. \cite{HubbleTensionReview}. It is thus safer to judge the theory in the context of the far more varied and unbiased distance moduli data, widely taken as the basis for late-time predictions \cite{HubbleTensionReview}. Nevertheless, it is interesting to note how the $n=10$ model provides an estimate for $H_0$ within $2\sigma$ of the reference late value when fitting the CC+BAO data, which reinforces the motivation for considering combinations of different exponents of $R$ in $f_2$, as discussed at the end of Section \ref{AccelerationSection}.\par

\section{Late-time Acceleration}\label{AccelerationSection}
Besides providing a natural agreement between direct and indirect measurements of the Hubble constant, we also consider the resulting late-time acceleration of the Universe, as revealed by recent data \cite{AcceleratedExpansion}. Indeed, as we have postulated a model with no cosmological constant, any deviation from the standard matter-dominated value of $q=1/2$, which describes a constantly decelerating expansion, must follow from the modifications introduced by the NMC model. However, as the main purpose of this work was to resolve the Hubble tension, we fix the corresponding value of $R_0$ for each $n$ and analyse the compatibility of these very models with observations of late-time acceleration. \par 
Throughout this section, we will continue to consider the values $n=4,10$ as references for the behaviour of the spectrum of values for $n$. As will become clear, both of these models have a smooth transition from the matter-dominated regime ($q=1/2$) at large $z$ to different asymptotic regimes as $z\rightarrow-1$. These asymptotic values seem to be in agreement with the results of Ref. \cite{NMCAcceleratedExpansion}, where a power-law form of the scale factor was considered in the limit of weak/strong matter-curvature coupling. As expected from the rapid increase of the magnitude of $F_2\tilde\rho$ as the curvature becomes increasingly smaller, our values agree with the strong coupling regime. This provides further motivation for the analysis of a power-law form for the expansion of the Universe in the future of the NMC gravity model, as conducted in the same paper \cite{NMCAcceleratedExpansion}. \par
\begin{figure} [h!]
    \centering
    \includegraphics[width=0.55\linewidth]{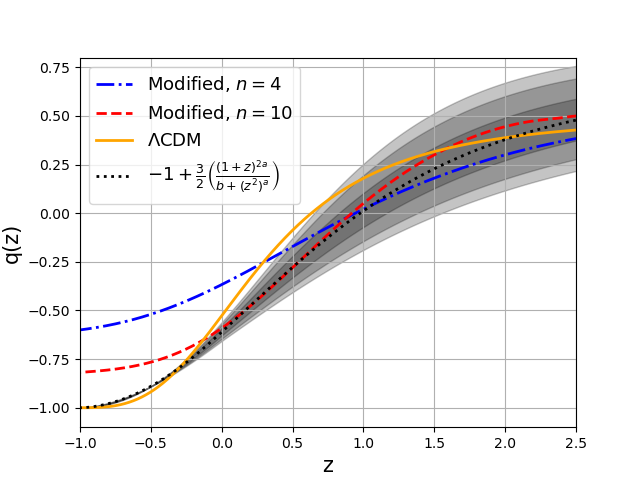}
    \caption{Comparison of observationally fitted parametrisation of $q(z)$ \cite{AcceleratedExpansion2} with the results for the Hubble tension-free NMC model. The shaded regions show the $1\sigma, \ 2\sigma$ and $3\sigma$ errors of the empirical fit. Also presented is the prediction given by the $\Lambda$CDM model.}
    \label{qParametrisation1}
\end{figure}

We consider two different empirical fittings for $q(z)$. The first, proposed in Ref. \cite{AcceleratedExpansion2} as the best of 3 parametrizations, is given by
\begin{equation}
    q(z)=-1+\frac{3}{2}\left(\frac{(1+z)^{2a}}{b+(z^2)^a}\right)
\end{equation}
where $a=0.855\pm0.034$ and $b=3.85\pm0.19$, and is shown in Figure \ref{qParametrisation1}. The fit for this model was obtained using the previously mentioned CC and BAO data \cite{BAO_BOSSData,CosmicChronometers}, together with the also mentioned distance moduli \cite{SHOESData,PantheonData}. Note that this function approaches $q=-1$ as we move towards the future ($z=-1$), similarly to the behaviour of standard $\Lambda$CDM. Interestingly, while the $n=4$ model tends to stray from the empirical fitting region near $z=0$, the $n=10$ model seems to agree with this fit, being within $1\sigma$ of the fitted model up until the same point. All of these agree on a transition redshift $z_T\approx0.9$, while distinctly disagreeing with the transition redshift of the standard model, this being typically around $z=0.65$. However, the modified models predict a weaker asymptotic acceleration phase in the far future, while the asymptotic value of the deceleration for the empirical fit follows $\Lambda$CDM  and is thus fixed at -1. In fact, even though our agreement between the empirical and NMC models could be expected from their agreement with the same observational data, we should note that each NMC model was merely chosen to remove the Hubble tension, with their respective consistency with other observational data being a consequence and not a requirement. \par

\begin{figure}[h!]
    \centering
    \includegraphics[width=0.55\linewidth]{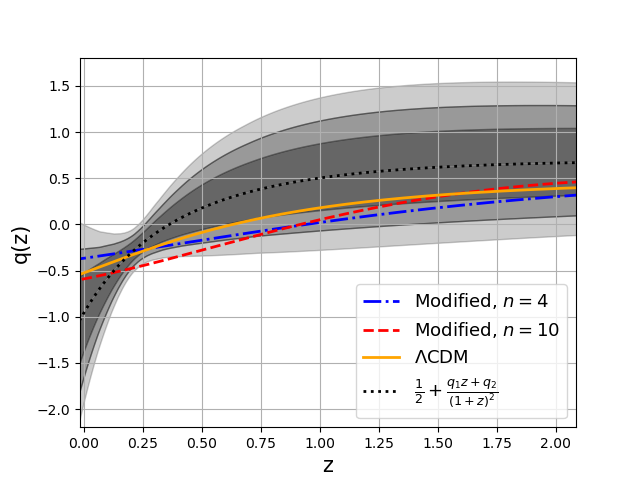}
    \caption{Comparison of observationally fitted parametrisation of $q(z)$ \cite{AcceleratedExpansion4} with the results for the Hubble tension-free NMC model. The shaded regions show the $1\sigma, \ 2\sigma$ and $3\sigma$ errors of the empirical fit. Also presented is the prediction given by the $\Lambda$CDM model.}
    \label{qParametrisation2}
\end{figure}

The second parametrisation, presented in Ref. \cite{AcceleratedExpansion4}, is given by
\begin{equation}
    q(z)=\frac{1}{2}+\frac{q_1 z+q_2}{(1+z)^2},
\end{equation}
where $q_1=1.47^{+1.89}_{-1.82}$ and $q_2=-1.46\pm0.43$, and is shown in Figure \ref{qParametrisation2}. This fit was obtained from the data for 182 gold type Ia supernovae \cite{GoldSNIa}, which determined the distance of these astrophysical events for several redshifts around $z=1$, analogously to the Pantheon+SH0ES data taken in the formerly discussed parametrisation and in this work in Section \ref{MethodSection}. In this case, the $n=4$ model is the one which appears to be within an error of $(1-2)\sigma$ of the empirical form of $q(z)$, with its $n=10$ counterpart providing better agreement at smaller redshifts. Additionally, this parametrisation exhibits a much lower acceleration transition of $z_T\approx0.35$, with the $1\sigma$ region ranging from $z=0.3$ to the approximate $\Lambda$CDM value of $z=0.65$, placing the transition of the modified models only within 2$\sigma$. However, the parametrisation presents a divergence at $z=-1$, leading to an increasingly accelerated expansion of the Universe as we move past the present stage. It is thus possible to consider this as a pathological solution for $q(z)$, which relatively lowers its importance when compared with the formerly considered example.

\subsection{More complex combinations}
Considering the different compatibilities between lower and higher values of $n$ and observational constraints, we are led to consider the possibility of using linear combinations of inverse powers of $R$. Indeed, functions of the form
\begin{equation}
    f_2(R)=\sum_{n=1}^{\infty}\left(\frac{R_n}{R}\right)^n
\end{equation}
have already been examined in the literature and provide a more detailed description of phenomena such as galaxy rotation curves \cite{NMCDarkMatter,NMCDarkMatter2}. A more general power-law expansion with various integer powers of $n$ is not just able to capture the behaviour of NMC models at different scales, but also provides an alternative template to consider more complex functions of $R$.\par

Naturally, one should not expect that the presence of two linearly combined terms in $f_2$ translates into a linear effect in the evolution of $H(z)$ due to the non-linearity of the theory. Nonetheless, the possibility of better understanding the behaviour of various sources of cosmological data observed at different epochs of the late Universe strongly motivates such considerations. Unfortunately, this leads to forms of $F_2$ that cannot be directly inverted to give the value of $R$ at each point in the simulation, forcing us to resort to root-finding functions at each step of our method. This increases the error in the already sensitive numerical evolution of the system, which then leads to severe instabilities and meaningless results. The analysis of these linear combinations in this context is thus left as the topic of future research. 

\section{Conclusions}\label{ConclusionSection}
In this work, we have applied a modified theory of gravity which nonminimally couples curvature and matter to the Hubble tension problem. In order to do that, we started by taking data from the CMB as accurate and hypothesising that the differences in measurements arise from the subsequent model-dependent evolution of the Universe. To ensure concordance between our model and the CMB data, we take the simple form of $f_2=(R_0/R)^n$, where $n>0$ is some positive exponent, leading to a decoupling of matter and curvature at high redshifts. By numerically evolving the modified field equations, we find the necessary values of $R_0$ for each $n$ that evolve the CMB observables to their measured late-time correspondents, therefore eliminating the Hubble tension. \par
The functional form of $H(z)$ was also tested against observational data from Baryon Acoustic Oscillations, Cosmic Chronometers and the observed distance moduli of supernovae. We considered $n=4,10$ as specific examples of the behaviour of the NMC model and found that $n=10$ is in better agreement with some of the BAO data, while $n=4$ seems to better fit the supernovae distance data \cite{NMCAcceleratedExpansion}. Alternatively, we also fit the model directly to the same observational data and obtained predictions for $H_0$ from the model's evolution. When considering recent distance moduli data from the Pantheon+SH0ES collaboration \cite{PantheonData,SHOESData}, we have shown that the best fit of the $n=4$ model predicts a value for $H_0$ that is within error of the reference value from the R20 measurement \cite{R20Cepheids}, thus effectively bridging the gap between CMB data and late-time observations. \par
Additionally, we have assumed no cosmological constant in our model, allowing us to investigate if the same form of $f_2$ can generate late accelerated expansion, similarly to what was done in Ref. \cite{NMCAcceleratedExpansion}. By comparing with empirically motivated parametrisations of the deceleration parameter, we have confirmed that late-time acceleration can be recreated by the presence of a non-minimal coupling between matter and curvature. Both the $n=4$ and $n=10$ models yield a present value $q_0<0$, along with similar acceleration transition redshifts $z_T\approx0.9$, leading to the mimicking of the effects of a cosmological constant. This result is non-trivial and provides further proof of the versatility of the NMC model, which adequately competes with current models in matching cosmological behaviour at low redshifts even when just burdened by the initially proposed challenge of removing the Hubble tension.  \par
Due to the different characteristics of the various values of $n$ considered here, the possibility of more complex combinations of exponents in $f_2$ was raised, with numerical stability issues impeding the development of testable results. Nevertheless, this possibility of capturing cosmological behaviours at different scales is promising in providing a better match between theory and data. We should note that patching the Hubble tension and reproducing the accelerated expansion of the Universe at late times are a small subset of the phenomena that can be explained through the nonminimally coupled theory of gravity, as shown in many examples in the literature \cite{NMCAcceleratedExpansion,NMCFriedmann,NMCDarkMatter,NMCInflation,NMCSolarSystem,NMCCosmologicalPerturbations}. The NMC theory considered in this paper is therefore incomplete without all the terms in the power expansion of $f_2$, which would dominate at different scales. Such an expansion would have manifestly different behaviours in different scenarios, allowing for the simultaneous explanation of effects such as dark matter ($f_2\propto R^{-1},R^{-1/3})$ \cite{NMCDarkMatter,NMCDarkMatter2}, dark energy ($f_2\propto R^{-4},R^{-10}$) \cite{NMCAcceleratedExpansion} and inflation ($f_2\propto R,R^3$) \cite{NMCInflation2,NMCInflation3}, among others. This combination of terms can be interpreted as the effective form of a putative full expansion of a more general nonminimal coupling that provides a more consistent description of gravity. Naturally, this interpretation further depends on arguments about the origin and the emergence of this non-minimal coupling, which remains a promising topic for future research in the field.

\begin{acknowledgments}
The work of one of us (O.B.) is partially supported by FCT (Fundação para a Ciência e Tecnologia, Portugal) through the project CERN/FIS-PAR/0027/2021, with DOI identifier  10.54499/CERN/FIS-PAR/0027/2021.
\end{acknowledgments}

\bibliographystyle{JHEP}
\bibliography{References.bib}

\end{document}